\documentstyle[psfig,preprint,prl,aps]{revtex}
\newcommand{\be}{\begin{equation}}
\newcommand{\ee}{\end{equation}}
\newcommand{\bea}{\begin{eqnarray}}
\newcommand{\eea}{\end{eqnarray}}
\newcommand{\bsub}{\begin{subequations}}
\newcommand{\esub}{\end{subequations}}
\def\vev{v.e.v.}
\def\gsim{~{\rlap{\lower 3.5pt\hbox{$\mathchar\sim$}}\raise 1pt\hbox{$>$}}\,}
\def\lsim{~{\rlap{\lower 3.5pt\hbox{$\mathchar\sim$}}\raise 1pt\hbox{$<$}}\,}
\def\lh{\lambda_H}
\def\lk{\lambda_K}
\def\ln{\lambda_N}
\def\mg{m_{3/2}}
\def\MHs#1{M_{H_#1}^2}
\def\MSs{M_S^2}
\def\PRD#1#2#3{Phys. Rev. {\bf D#1}, #3 (19#2)}
\def\NPB#1#2#3{Nucl. Phys. {\bf B#1}, #3 (19#2)}
\def\PTP#1#2#3{Prog. Theor. Phys. {\bf #1}, #3 (19#2)}

\def\PLB#1#2#3{Phys. Lett. {\bf B#1}, #3 (19#2)}
\def\PRL#1#2#3{Phys. Rev. Lett. {\bf #1}, #3 (19#2)}
\def\PRep#1#2#3{Phys. Rep. {\bf #1}, #3 (19#2)}
\begin{document}
\draft
\preprint{OCHA-PP-137} 

\title{ 
A Supersymmetric Model with an Extra U(1) Gauge Symmetry
}
\author{
Mayumi Aoki
}
\address{
 Graduate School of Humanities and Sciences, Ochanomizu University  \\
Otsuka 2-1-1, Bunkyo-ku, Tokyo 112-8610, Japan  \\
}
\author{
Noriyuki Oshimo
}
\address{
 Department of Physics, Ochanomizu University  \\
Otsuka 2-1-1, Bunkyo-ku, Tokyo 112-8610, Japan  \\
}
\date{\today}
\maketitle
\begin{abstract}

     In the standard model the proton is protected from decay 
naturally by gauge symmetries, whereas in the ordinary minimal 
supersymmetric standard model an ad hoc discrete symmetry 
is imposed for the proton stability.  
We present a new supersymmetric model in which 
the proton decay is forbidden by an extra U(1) gauge symmetry.
Particle contents are necessarily increased to be free from anomalies, 
incorporating right-handed neutrinos.  
Both Dirac and Majorana masses are generated for neutrinos, 
yielding non-vanishing but small masses.
The superpotential consists only of trilinear couplings
and the mass parameter $\mu$ of the minimal model is induced 
by spontaneous breaking of the U(1) symmetry.
\end{abstract}

\narrowtext

     Extending the standard model (SM) by supersymmetry \cite{nilles}, 
though considered to be promising for physics above the electroweak 
energy scale, is confronted with a problem of the proton stability.  
Among the possible interactions allowed by the gauge symmetries 
of the SM and renormalizability, those which violate baryon-number 
or lepton-number conservation are included, 
leading to an unacceptably fast decay of the proton.  
For forbidding these interactions, 
an ad hoc discrete symmetry is usually imposed on   
supersymmetric models through $R$ parity, 
as on the minimal supersymmetric standard model (MSSM).  
On the other hand, in the SM, baryon and lepton numbers are 
conserved merely as a consequence of gauge symmetries.   
Some more fundamental reason for the proton stability 
should exist also in the supersymmetric standard model.  

     Neutrino masses also pose a potential problem for the MSSM.
Experiments for atmospheric and solar neutrinos, 
such as at the Super-Kamiokande \cite{kamiokande}, suggest non-vanishing 
masses for the neutrinos.  
Theoretically, Yukawa couplings 
which generate their Dirac masses can be included naively 
by incorporating the superfields of right-handed neutrinos.  
However, these superfileds are inert for the transformations of the SM 
gauge groups and thus their raison d'etre is vague.  
Furthermore, if the extreme lightness of the neutrinos is explained by 
large Majorana masses for their right-handed components, 
new mass parameters of unknown origin have to be introduced.  

     Another potential problem is raised by the linear coupling of 
the Higgs superfields contained in the superpotential of the MSSM.  
Since the mass parameter $\mu$ of this coupling enters into 
the scalar potential which determines SU(2)$\times$U(1) symmetry breaking, 
its magnitude should be of the electroweak scale.  
Assuming the model coupled to $N=1$ supergravity, 
the other mass parameters in the scalar potential are traced back to 
supersymmetry-soft-breaking terms and thus related to 
the gravitino mass which can be taken as the electroweak scale.  
On the other hand, various mechanisms have been proposed to 
generate the $\mu$ parameter, although its origin is still 
controversial \cite{kim}.  

     Aiming at providing natural and consistent solutions for 
the above problems, in this letter, 
we propose a new supersymmetric standard model, 
based on the gauge symmetries with an extra U(1) group and 
$N=1$ supergravity.  
It is plausible that the proton stability is guaranteed by 
a gauge symmetry.  
In particular, one of the simplest possibilities is 
by U(1) \cite{weinberg,hall,oshimo,font,chamseddine}.   
We present a U(1) gauge symmetry which conserves baryon number 
but allows lepton-number violation, sufficient for the proton stability.   
Particle contents are necessarily 
increased in order to be free from gauge and trace anomalies, 
and then incorporate right-handed neutrinos.  
Also included is a scalar field
which has a non-vanishing vacuum expectation value (\vev) and breaks  
the U(1) gauge symmetry.  
This \vev\ induces Majorana masses for the right-handed neutrinos and 
an effective $\mu$ parameter for the Higgs linear coupling.  
The U(1) gauge symmetry predicts a new neutral gauge boson, 
for which non-trivial constraints have been obtained from experiments. 
We show that these constraints can be satisfied without fine-tuning 
much model parameters.   
A natural scale of this model becomes of order 1 TeV, 
which could account for the smallness of 
the electric dipole moments (EDMs) of the neutron and the electron, 
another problem in general supersymmetric models.   
 
     The model consists of the left-handed chiral superfields 
listed in Table \ref{particles},  
where shown are their quantum numbers under  
SU(3), SU(2), U(1), and U$'$(1) gauge transformations.  
The extra U(1) group is denoted by U$'$(1).  
The index $i$ stands for the generation.  
The U$'$(1) charges have been normalized as Tr($Y^2$)=Tr($Y'^2$), 
$Y$ and $Y'$ being respectively hypercharge and U$'$(1) charge generators.  
The generators $Y$ and $Y'$ are orthogonal, Tr($YY'$)=0.   
The gauge anomalies and the trace anomalies  
are canceled in each generation.  
New superfields which are not contained in the MSSM are 
SU(3) triplets $K^i$ and $K^{ci}$ and 
SU(3)$\times$SU(2)$\times$U(1) singlets $N^{ci}$ and $S^i$.  
The superfields for right-handed neutrinos are denoted by $N^{ci}$.  
In addition, there exist SU(2) doublets $H_1^i$ and $H_2^i$ 
in each generation, some or all of which assume 
SU(2)$\times$U(1) symmetry breaking as Higgs fields.  
In respect of the quantum numbers for the SM gauge groups, 
the particle contents resemble those of the extra-U(1) model 
based on the fundamental representation of the E$_6$ group. 
The difference is in hypercharge assignment to  
the new colored superfields, 
which is necessary to protecting the proton from decay by  
introducing an extra U(1) symmetry \cite{oshimo,font}.   
In such an E$_6$ model an additional discrete symmetry  
has to be imposed for forbidding the proton decay.   

     The superpotential is given by
\begin{eqnarray}
W &=& \eta_d^{ijk} H_1^iQ^jD^{ck} + \eta_u^{ijk}H_2^iQ^jU^{ck} 
 + \eta_e^{ijk} H_1^iL^jE^{ck} + \eta_\nu^{ijk} H_2^iL^jN^{ck}   \nonumber \\
&& + \lambda_N^{ijk}S^iN^{cj}N^{ck} + \lambda_H^{ijk}S^iH_1^jH_2^k 
  + \lambda_K^{ijk}S^iK^jK^{ck}, 
\label{superpotential}
\end{eqnarray}
where contraction of group indices is understood.  
This superpotential contains all the terms consistent 
with gauge symmetries and renormalizability.  
The interactions which would violate baryon- or lepton-number 
conservation in the MSSM, 
i.e. $D^cD^cU^c$, $LQD^c$, $LLE^c$, $H_1H_1E^c$, and $LH_2$, are not allowed.  
In fact, baryon number is conserved while lepton number is not.  
The lowest dimension operators for baryon-number 
violation are given by the D terms of $QQU^{c*}E^{c*}$, $QQD^{c*}N^{c*}$, 
and $QU^{c*}D^{c*}L$, which have dimension 6.  
Therefore, the proton decay is suppressed at least by a huge mass to 
the second power.  
If this mass is of order the 
energy scale of grand unified theories (GUTs) or larger, 
the proton becomes adequately stable.   
The couplings in Eq. (\ref{superpotential}) 
are all cubic, and there is no dimensionful parameter. 
The terms $H_1QD^c$, $H_2QU^c$, $H_1LE^c$, and $H_2LN^c$ yield 
Dirac masses for quarks and leptons including neutrinos.  
Through a non-vanishing \vev\ of the scalar component of $S$, 
the term $SN^{c}N^{c}$ generates a Majorana mass for the 
right-handed neutrino and the term $SH_1H_2$ serves as the linear 
coupling of the Higgs superfields in the MSSM.  
The term $SKK^{c}$ induces a mass for the fermion 
components of $K$ and $K^{c}$.  

     The model is coupled to $N=1$ supergravity, 
which is spontaneously broken in a hidden sector at the Planck mass scale.  
Below the GUT scale 
the Lagrangian of the observable sector consists of a supersymmetric 
part and a supersymmetry-soft-breaking part prescribed 
by gauge symmetries and superpotential.  
The soft-breaking part contains mass terms for scalar bosons and 
gauge fermions, and trilinear couplings for scalar bosons.    

     We now examine the vacuum structure of this model.   
The SU(2)$\times$U(1)$\times$U$'$(1) gauge symmetry can be  
spontaneously broken by non-vanishing \vev s for the scalar components 
of $H_1^i$, $H_2^i$, and $S^i$.  
Since it is too complicated to discuss the vacuum with all 
of them being taken into account, we assume only one set of them to 
have non-vanishing \vev s, for simplicity. 
The scalar potential is then given by 
\begin{eqnarray}
V &=& \frac{1}{8}g_2^2\left( |H_1|^2+|H_2|^2\right) ^2
     +\frac{1}{8}g_1^2\left( |H_1|^2-|H_2|^2\right) ^2 \nonumber  \\
    & & +\frac{1}{72}g'^2\left( 4|H_1|^2+|H_2|^2
                       -5|S|^2\right) ^2  \nonumber  \\
    & & -\left( \frac{1}{2}g_2^2-|\lh|^2\right) |H_1H_2|^2 
              +|\lh|^2\left( |H_1|^2+|H_2|^2\right)|S|^2  \nonumber \\
    & & +\left( B_H\lh\mg SH_1H_2+{\rm H.c.}\right)
                      +\MHs1|H_1|^2+\MHs2|H_2|^2+\MSs|S|^2,  
\label{potential}
\end{eqnarray}
where $\mg$ denotes the gravitino mass, $B_H$ being a dimensionless 
constant, and $\MHs1$, $\MHs2$, and $\MSs$ represent mass-squared 
parameters.     
The gauge coupling constants for SU(2), U(1), and U$'$(1) are denoted 
by $g_2$, $g_1$, and $g'$, respectively.   
We have adopted the same notation for the superfields and 
their scalar components.  
Differently from the MSSM, there is no D-flat direction where 
quartic couplings of Higgs fields are absent, and the potential has 
a stable minimum irrespectively of the supersymmetry-soft-breaking terms.  
If the condition $g_2^2>2|\lh|^2$ is satisfied, electric charge is conserved.  
Redefining the global phases of the fields so as to 
give $B_H\lh=-|B_H\lh|$, the \vev s 
$v_1$, $v_2$, and $v_s$ of the neutral components of $H_1$, $H_2$, 
and $S$, respectively, become real and non-negative.   
These values are determined by extremum conditions  
$\partial V/\partial v_1=0$, $\partial V/\partial v_2=0$, 
and $\partial V/\partial v_s=0$.  
It turns out that the solution of these simultaneous equations 
with $v_1$, $v_2$, and $v_s$ all non-vanishing is unique, if exists.  
The true vacuum is either at such a point or 
a point on the boundary $v_1v_2v_s=0$, which can be identified 
by comparing the potential energies of those points.  

     On the \vev s, there exist experimental constraints.  
The $W$ boson mass has been measured precisely.  
Furthermore,  there appear two massive neutral gauge bosons 
$Z_1$ and $Z_2$ ($M_{Z_1}<M_{Z_2}$) as mass eigenstates of the $Z$ 
boson for SU(2)$\times$U(1) and the $Z'$ boson for U$'$(1).   
The measured mass of $Z$ for the SM should be taken as the mass of $Z_1$;   
the lower bound on the mass of $Z_2$ is given by 
$M_{Z_2}\gsim 600$ GeV \cite{cdf};   
and the mixing between $Z$ and $Z'$ should be sufficiently small, 
roughly given by $A_{ZZ'}^2/A_{ZZ}A_{Z'Z'}(\equiv R)\lsim 10^{-3}$,  
where $A_{ZZ}$, $A_{Z'Z'}$, and $A_{ZZ'}$ represent the 
elements of the mass-squared matrix $A$ for $Z$ and $Z'$,  
according to analyses of various measurements for 
electroweak parameters \cite{cho}.  
These constraints require in some degree non-trivial differences 
of scale between the \vev s.  

     The scalar potential in Eq. (\ref{potential}) gives a plausible 
vacuum under certain ranges of parameter values, 
which is seen by numerical analyses.  
The above constraints on the \vev s can be generally satisfied 
if the typical scale of the mass parameters in the potential 
is larger than 1 TeV.  
However, as the mass scale increases, more fine-tuning 
becomes inevitable to obtain the hierarchy of the \vev s.  
Therefore, it would be natural to consider the mass scale to be 
of order 1 TeV. 
We present an example of the vacuum in Tables \ref{parameters} 
and \ref{masses}.  
In Table \ref{parameters} the values of  
the parameters in the potential are shown, where   
the gauge coupling constant for U$'$(1) is taken for $g'=g_1$.  
The \vev s, $M_{Z_2}$, $R$, and the masses of the physical Higgs bosons 
are shown in Table \ref{masses}, where $H^0$, $A^0$, and $H^\pm$ stand for 
the neutral scalar, neutral pseudoscalar, and charged Higgs bosons, 
respectively.  
The Higgs boson masses have been calculated, assuming for definiteness 
that the Higgs fields $H_1$, $H_2$, $S$ form mass eigenstates 
by themselves without mixing with the other fields 
of $H_1^i$, $H_2^i$, and $S^i$.  
In this case the neutral Higgs bosons do not mediate interactions 
of flavor-changing neutral current, thus causing no effect on 
$K^0$-$\bar K^0$ mixing.  
As in other supersymmetric models, one Higgs boson is light.  
This mass could increase non-negligibly 
if one-loop corrections are incorporated.  
If the \vev s are set for $v_2/v_1=2$, the mixing parameter $R$ vanishes .  

     In Tables \ref{parameters} and \ref{masses},  
the dimensionless parameters have reasonable values,  
and the differences between the mass parameters 
are at most of one order of magnitude, suggesting that only 
mild fine-tuning is required.  
Scalar fields are considered to have supersymmetry-soft-breaking 
masses of order the gravitino mass $\mg$ at around the GUT scale.  
Then, $\MHs2$ and $\MSs$ receive large negative contributions 
through quantum corrections at the electroweak scale,  
owing to the couplings $H_2Q^3U^{c3}$ and $SK^jK^{ck}$ with 
large coefficients.  
Therefore, a small value for $\MHs2$ and a negative value for $\MSs$,  
as shown in this example, are likely to occur \cite{langacker}.  
On the other hand, quantum corrections to $\MHs1$ are 
not so large, and its value should remain around $\mg^2$.  
Similarly, the scalar particles other than the Higgs bosons have masses 
of order the gravitino mass.     
Although the masses-squared for the scalar components of $K^i$ and $K^{ci}$ 
receive non-negligible negative contributions from the D-term of U$'$(1),  
the positive contributions from the soft-breaking terms 
dominate over, keeping SU(3) symmetry unbroken.  
The scalar components of $N^{ci}$ also do not have non-vanishing \vev s.  

     The neutrino masses and the effective $\mu$ parameter are generated 
in realistic ranges.  
For $\ln=0.1$, taking a neutrino Dirac mass for the same as 
the electron mass, the lighter mass eigenvalue becomes about 0.5 eV, 
which varies in proportion to the square of the Dirac mass.    
The effective $\mu$ parameter is  given by $\lh v_s/\sqrt{2}$, 
leading to $|\mu|\approx 240$ GeV.  
Assuming that the gauge fermions for SU(2), U(1), and U$'$(1) 
receive masses of order 100 GeV from the soft-breaking terms, 
the masses of the lighter chargino and the lightest neutralino  
become of order 100 GeV.  

     The numerical analyses provide, as a byproduct, an explanation for 
a problem on the EDMs of the neutron and the electron.  
If one assumes the squark and slepton masses of order 100 GeV and 
the $CP$-violating phases intrinsic in supersymmetric models unsuppressed,  
these EDMs are predicted to be much larger than their experimental 
upper bounds.    
However, in this model, the lower bound on the $Z_2$ boson mass 
implies that a natural scale for soft-breaking masses of 
scalar fields are of order 1 TeV.   
Consequently, the squarks and sleptons have masses  
of this order of magnitude.  
Then, the constraints from the EDMs become theoretically amenable and the 
$CP$-violating phases need not be fine-tuned very small \cite{kizukuri}.  
If these phases are not suppressed, 
sizable $CP$ violation is expected to occur in some reactions 
at the energy scale of order 1 TeV. 
In particular, since lepton number is also violated, 
certain reactions could involve both $CP$ violation and lepton-number 
violation.  
For instance, the Higgs boson which is mainly composed 
of $S$ can decay into both $\nu_R\nu_R$ and $\bar\nu_R\bar\nu_R$, whose 
branching ratios could have different values.   
Such reactions may lead to a non-vanishing lepton 
number in the universe at around its electroweak phase transition.  
The baryon asymmetry of the universe could then be generated  
by converting the net lepton number through the sphaleron process.   

     In this model, a Dirac fermion which is composed of the fermion 
components of $K$ and $K^c$ becomes stable, having both color 
and electric charges. 
Its mass is given by $\lk v_s/\sqrt{2}$, which is of order 
100 GeV $-$ 1 TeV.    
The particle could thus be detectable in near-future experiments.  
On the other hand, such a stable particle may also be explored  
by non-accelerator experiments, e.g. search for anomalous 
nuclei in seawater, provided that its relic density in the 
universe is not very small.  
In fact, a purely perturbative calculation for the pair 
annihilation of this particle leads to a density which 
may be inconsistent with constraints from such experiments.  
However, the annihilation cross section of a colored 
particle could be extremely enhanced by non-perturbative effects, 
which may render the density too small to be detected.   
Since these effects are not yet understood well quantitatively, 
large uncertainties of as much as ten orders of 
magnitude could emerge in the calculation of the relic 
density \cite{baer}, making a definite prediction difficult.   
In addition, some cosmological reasons, such as possible 
low-energy inflation, could dilute the relic density well below the 
detectable level.  
Therefore, definite constraints on the stable particle should 
come only from accelerator experiments.    

     In summary, we have presented a new supersymmetric standard model
based on SU(3)$\times$SU(2)$\times$U(1)$\times$U$'$(1) gauge 
symmetry and $N=1$ supergravity.  
In this model, the proton is stable by gauge symmetries 
without invoking a further symmetry.  
The right-handed neutrinos are introduced as fields 
which are to cancel anomalies.  
After U$'$(1) symmetry is spontaneously broken, 
large Majorana masses are induced for the right-handed neutrinos, 
leading naturally to light neutrinos consistent with experiments.  
The effective $\mu$ parameter is also generated by the symmetry breaking.  
A natural energy scale of this model is of order 1 TeV, 
which does not require excessive fine-tuning of parameters 
for electroweak symmetry breaking.  
The EDMs of the neutron and the electron can also be accommodated 
without fine-tuning $CP$-violating phases.  
     
\smallskip

The authors thank G.C. Cho for discussions on 
the extra U(1) gauge boson in the E$_6$ model.  
One of the authors (M.A.) acknowledges the Japan 
Society for the Promotion of Science for financial support.   
The work of M.A. is supported in part by 
the Grant-in-Aid for Scientific 
Research from the Ministry of Education, Science and Culture, Japan.  

%

\newpage
%
%
%
%
%
%
%
%
%
%
%

\begin{table}
\caption{Particle contents and their quantum numbers.
        }
\label{particles}
\begin{tabular}{l l c r r}
   &  SU(3) & SU(2) & U(1) & U$'$(1) \\  
\hline
$Q^i$    & 3   & 2 & $\frac{1}{6}$  & $\frac{1}{12}$  \\ 
$U^{ci}$    & $3^*$ & 1 & $-\frac{2}{3}$ & $\frac{1}{12}$  \\ 
$D^{ci}$    & $3^*$ & 1 & $\frac{1}{3}$  & $\frac{7}{12}$  \\     
$L^i$    & 1   & 2 & $-\frac{1}{2}$ & $\frac{7}{12}$  \\ 
$N^{ci}$    & 1   & 1 &      0         & $-\frac{5}{12}$ \\
$E^{ci}$    & 1   & 1 &      1         & $\frac{1}{12}$  \\
$H_1^i$  & 1   & 2 & $-\frac{1}{2}$ & $-\frac{2}{3}$  \\
$H_2^i$  & 1   & 2 & $\frac{1}{2}$  & $-\frac{1}{6}$  \\
$S^i$    & 1   & 1 &      0         & $\frac{5}{6}$   \\
$K^i$    & 3   & 1 & $\frac{1}{3}$  & $-\frac{2}{3}$  \\
$K^{ci}$ & $3^*$ & 1 & $-\frac{1}{3}$ & $-\frac{1}{6}$  \\ 
\end{tabular}
\end{table}
\begin{table}
\caption{Parameter values.}
\label{parameters}
\begin{tabular}{c c c c c c}
$g'$ & $|\lh|$ & $|B_H\mg|$ & $\MHs1$ & $\MHs2$ & $\MSs$      \\ 
\hline
 0.36 & 0.10 & 1.0 TeV & (1.2 TeV)$^2$ & (0.29 TeV)$^2$ 
                            & $-$(0.71 TeV)$^2$  \\
\end{tabular}
\end{table}
\begin{table}
\caption{The \vev s and masses.}
\label{masses}
\begin{tabular}{c c c c c }
$v_2/v_1$ & $v_s$ & $M_{Z_2}$ & $R$ &   \\ 
\hline
5.0 & 3.4 TeV & 1.0 TeV & 1.4$\times 10^{-4}$ &  \\
\hline
$M_{H^0}$ & & & $M_{A^0}$ & $M_{H^\pm}$  \\
\hline
85 GeV & 1.0 TeV & 1.1 TeV & 1.1 TeV & 1.1 TeV \\
\end{tabular}
\end{table}

\end{document}